\documentclass[prc,twocolumn,superscriptaddress,showpacs,preprintnumbers,amsmath,amssymb]{revtex4}
\usepackage[dvips]{graphicx,color}
\usepackage{dcolumn,bm}
\def\m@thcombine#1#2{%
  \setbox0=\hbox{$#1$}
  \setbox1=\hbox{$#2$}
  \ifdim\wd0>\wd1
    \setbox0=\hbox to\wd1{\hss\box0\hss}
  \else
    \setbox1=\hbox to\wd0{\hss\box1\hss}
  \fi
  \mathop{\vcenter{
    \offinterlineskip\box0\box1}}}
\def\lesim{\m@thcombine<\sim}
\def\gesim{\m@thcombine>\sim}
\def\vr{\mbox{\boldmath$r$}}
\newcommand{\vect}[1]{\mbox{\boldmath$#1$}}
\def\+{\mbox{\unboldmath $+$}}
\def\-{\mbox{\unboldmath $-$}}
\def\={\mbox{\unboldmath $=$}}

\begin{document}
\title{Mean-field analysis of ground state and low-lying electric dipole strength
in $^{22}$C}
\author{T. Inakura}
\affiliation{Department of Physics, Graduate School of Science, Chiba University, Chiba, 263-8522, Japan}
\author{W. Horiuchi}
\affiliation{Department of Physics, Hokkaido University, Sapporo 060-0810, Japan}
\author{Y. Suzuki}
\affiliation{Department of Physics, Niigata University, Niigata 950-2181, Japan}
\affiliation{RIKEN Nishina Center, Wako, 351-0198, Japan}
\author{T. Nakatsukasa}
\affiliation{RIKEN Nishina Center, Wako, 351-0198, Japan}
\affiliation{Center for Computational Sciences, University of Tsukuba, Tsukuba 305-8571, Japan}

\begin{abstract}
Properties of neutron-rich $^{22}$C are studied using
the mean-field approach with Skyrme energy density functionals.
Its weak binding and large total reaction cross section,
which are suggested by recent experiments,
are simulated by modifying the central part of Skyrme potential.
Calculating $E1$ strength distribution by using the random-phase approximation,
we investigate developments of low-lying electric dipole ($E1$) strength 
and a contribution of core excitations of $^{20}$C.
As the neutron Fermi level approaches the zero energy threshold 
($\varepsilon_F \gesim -1$ MeV),
we find that the low-lying $E1$ strength exceeds the energy-weighted cluster
sum rule, which indicates an importance of the core
excitations with the $1d_{5/2}$ orbit.
\end{abstract}

\pacs{
21.10.Pc, 
25.20.-x, 
27.30.+t 
}
\maketitle

\section{Introduction}

The neutron drip-line nucleus $^{22}$C is the heaviest Borromean system 
that we have found so far. An early study for $^{22}$C was done by two of 
the present authors (W.H. and Y.S.)~\cite{Horiuchi06} with a three-body model
of $^{20}\mathrm{C}+n+n$ and predicted an $s$-wave dominance of $^{22}$C showing 
a large matter radius $r_\mathrm{rms}=$ 3.58 -- 3.74 fm which is comparable 
to those of medium mass nuclei. Recently Tanaka {\it et al.}~\cite{Tanaka10}
measured a very large total reaction cross section on a proton target and
a simplified three-body model analysis gives an empirical matter radius 
$5.4 \pm 0.9$ fm which is so large that is comparable to the radius of $^{208}$Pb. 
Gaudefroy {\it et al.}~\cite{Gaudefroy10} measured masses of several neutron-rich 
nuclei and found a very small two-neutron separation energy, $S_{2n}= -0.14 \pm 0.46$ 
MeV, of $^{22}$C. The small $S_{2n}$ implies the two-neutron halo structure having the large matter radius.

This large matter radius has attracted great attention. Since $^{22}$C has 
large uncertainly of the two-neutron separation energy and the property of 
an unbound $^{21}$C nucleus is not well known, several theoreticians try to 
constrain the binding energy from that empirical matter radius \cite{Yamashita11,Fortune12}.
The extended neutron orbits in the ground state affect the low-lying 
excitation of $^{22}$C. Large enhancement of the low-lying electric dipole ($E1$) strength 
was predicted in a three-body model with small two-neutron separation energy \cite{Ershov12}.

The low-lying $E1$ strength in medium-mass and heavy nuclei, which is often 
called pygmy dipole resonance (PDR), is of particular interest in relation 
with the properties of the neutron matter \cite{Carbone10}. However, this 
excitation mechanism is not understood well with regard to whether the mode 
is collective, single-particle excitations or not. Also it 
is interesting to ask a question of whether or not the enhancement of the 
low-lying $E1$ strength is universal in neutron-rich nuclei.

Though some theoretical works are devoted to understand the structure of 
$^{22}$C, all discussions are based on a three-body model which is often 
employed to describe light halo nuclei. The three-body model seems to also 
work well for $^{22}$C because the $s$-wave dominance, being associated with 
the $N=14$ subshell closure, is confirmed~\cite{Tanaka10}. On the other hand, 
as is pointed out in Refs.~\cite{Otsuka02,Hamamoto07}, the robustness of the 
$N=14$ subshell closure is weakened in very neutron-rich nuclei. Therefore a 
study without an assumption of the frozen $^{20}$C core is desired for deep 
understanding of the structure of $^{22}$C.

In this paper, we present a detailed analysis of the structure of 
$^{22}$C without assuming the $^{20}$C core through its low-lying $E1$ 
strength. We calculate the ground state properties and the low-lying $E1$ 
strength with the mean-field approach, namely, Hartree-Fock (HF) calculation 
and random-phase approximation (RPA) with the Skyrme density functional. The 
calculation is performed in a self-consistent manner. 

The manuscript is organized as follows. Section \ref{sec.model} reviews 
briefly the HF and RPA calculation. In Sec. \ref{sec.Result}, we analyze 
the ground state properties of $^{22}$C obtained with original Skyrme 
interactions. To reproduce the halo structure in $^{22}$C,
we search for the best 
parameters of the Skyrme interaction and tune them accordingly.
The validity of the interaction is tested by analyzing the total reaction
cross sections in comparison with the experimental ones. We discuss in 
detail the low-lying $E1$ strength and the excitation mechanism.
Conclusions are given in Sec. \ref{sec.Conclusion}.

\section{Models}
\label{sec.model}

We perform the HF calculation for $^{22}$C with the Skyrme interaction. 
The ground state is obtained by minimizing the following energy density 
functional~\cite{Vautherin72},
\begin{eqnarray}
E[\rho] = E_N + E_C - E_\mathrm{cm} .
\end{eqnarray}
For the ground state, the nuclear energy $E_N$ is given by a functional of the nucleon 
density $\rho_q(\vr)$, the kinetic density $\tau_q(\vr)$, the spin-orbit-current density 
$\nabla\cdot\vect{J}_q(\vr)$ ($q=n,\,p$).
The Coulomb energy $E_C$ among protons is 
a sum of direct and exchange parts. The exchange part is approximated by means of the 
Slater approximation, $\propto \int \mathrm{d}\vr \, \rho_p(\vr)^{4/3}$. 

Every single-particle wave function $\phi_i(\vr)$ is represented in the 
three-dimensional grid points with the adaptive Cartesian mesh~\cite{Nakatsukasa05}.
All the grid points inside the sphere of radius $R_\mathrm{box}=50$ fm are adopted in 
the model space. All the single-particle wave functions and potentials except for 
the Coulomb potential are assumed to vanish outside the sphere. For the calculation 
of the Coulomb potential, we follow the prescription in Ref.~\cite{Flocard78}. The 
differentiation is approximated by a finite difference with the nine-point 
formula. The ground state is constructed by the imaginary-time method~\cite{Davies80} 
with the constraints on the center-of-mass and the principal axes
\begin{eqnarray}
\langle x \rangle = \langle y \rangle = \langle z \rangle = 0 ,\quad
\langle xy \rangle = \langle yz \rangle = \langle zx \rangle = 0 .
\end{eqnarray}

On top of the ground state obtained by the Skyrme-HF, we calculate low-lying $E1$ 
strength using the RPA approach~\cite{Ring-Schuck}. We calculate the linear 
response for the $E1$ external field $V_\mathrm{ext}$ at a fixed complex energy 
$\omega = E + i \gamma /2$ by an iterative solver, the generalized conjugate 
residual method~\cite{Eisenstat}. The imaginary part of the energy is fixed at 
0.5 MeV, corresponding to smearing with $\gamma=1.0$ MeV. The calculation is done 
self-consistently with the Skyrme energy functional, including time-odd
densities. The residual field 
$\delta h \,(= (\partial^2 E / \partial \rho\partial \rho)\cdot \delta\rho)$
induced by $V_\mathrm{ext}$ contains all the terms including the time-odd 
components, the residual spin-orbit interaction,
and the residual Coulomb interaction. To facilitate an achievement of the self-consistency, 
we use the finite amplitude method (FAM) 
\cite{Nakatsukasa07,Inakura09,Inakura11,Avogadro11,Stoitsov11,Avogadro13,Liang13,Hinohara13}. 
The FAM allows us to evaluate the self-consistent residual fields as a 
finite difference, employing a computational code for the static mean-field 
Hamiltonian alone with a minor modification.

In the RPA, the transition density $\delta\rho$ 
at a complex energy $\omega$ is expressed,
with the forward and backward amplitudes, $X_i(\omega,\vr)$ and $Y_i(\omega,\vr)$, as
\begin{eqnarray}
\delta \rho(\omega,\vr) \equiv  -\frac{1}{\pi} \mathrm{Im} \sum_{i\in \mathrm{occ.}} \left\{ \phi^\ast_i(\vr)  X_i(\omega,\vr)  +  Y^\ast_i(\omega,\vr) \phi_i(\vr)  \right\} ,
\label{drho}
\end{eqnarray}
where $i$ runs over the occupied orbits
and the spin indices are omitted for simplicity.
In this article, we consider an $E1$ operator for the external field
\begin{equation}
D_z =\frac{Ne}{A} \sum^Z_{p=1} r_p Y_{10}(\Omega_p) -\frac{Ze}{A} \sum^N_{n=1} r_n Y_{10} (\Omega_n) ,
\label{D_E1}
\end{equation}
and similar operators for $D_x$ and $D_y$. 
The $E1$ strength for a real frequency $\omega=E$ is expressed by
\begin{eqnarray}
&& S(E;E1) \equiv \sum_n \left| \langle n | D | 0 \rangle \right|^2 \delta(E-E_n) \nonumber \\
&& \phantom{S(E;E1)} = -\frac{1}{\pi} \mathrm{Im} \sum_i \left\{ \langle \phi_i | D | X_i(\omega) \rangle + \langle Y_i(\omega)  | D| \phi_i \rangle \right\}, \nonumber \\
\label{dB/dw}
\end{eqnarray}
where $| n\rangle$ are energy eigenstates of the total system. 
For the complex energies $\omega$, the $E1$ strength becomes 
\begin{eqnarray}
&& S(E;E1) = \frac{\gamma}{2\pi} \sum_n \left\{ \frac{ | \langle n | D | 0 \rangle  |^2 } {\left( E - E_n \right)^2 + \left( \gamma /2 \right)^2} \right. \nonumber \\
&& \phantom{S(E1)= \frac{\gamma}{2\pi} \sum_n \{\quad } \left. - \frac{ | \langle n | D | 0 \rangle |^2 } {\left( E + E_n \right)^2 + \left( \gamma /2 \right)^2} \right\} .
\label{FAM_smearing}
\end{eqnarray}
The calculated strength is interpolated using the cubic spline function. 
The computer program employed in the present work has been developed 
previously~\cite{Nakatsukasa07,Inakura09,Inakura11}.

\section{Results and discussion}
\label{sec.Result}

\subsection{Ground state properties}

\begin{figure}[tb]
\begin{center}
\includegraphics[width=0.420\textwidth,keepaspectratio]{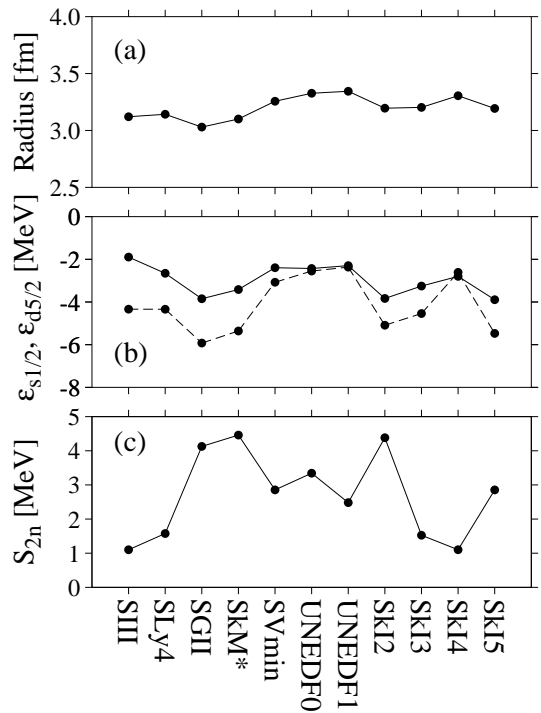}
\caption{
Ground state properties of $^{22}$C calculated with various Skyrme interactions: 
(a) rms matter radius, (b) neutron single-particle energies of $2s_{1/2}$ (solid) and $1d_{5/2}$ 
(dashed) orbits, and (c) two-neutron separation energy. 
}
\label{fig.data}
\end{center}
\end{figure}

Here we test the ground state properties calculated within the HF approximation. We adopt a variety of 
Skyrme functionals; SIII~\cite{SIII}, SLy4~\cite{SLy4}, SGII~\cite{SGII}, SkM$^*$~\cite{SkM*}, 
SVmin~\cite{SVmin}, UNEDF0, UNEDF1~\cite{UNEDF}, SkI2, SkI3, SkI4, and SkI5~\cite{SkI}. Figure 
\ref{fig.data} shows the calculated ground state properties of $^{22}$C. 
All these Skyrme interactions produce root-mean-square (rms) radii in the range of $r_\mathrm{rms} = 3.03 - 3.34$ 
fm, which is smaller than that obtained by the three-body 
model, 3.58 -- 3.74 fm~\cite{Horiuchi06} and the experimental value, $5.4 \pm 0.9$ fm. These small 
calculated radii are connected with the single-particle energy of the $2s_{1/2}$ orbit. As shown in 
Fig.~\ref{fig.data} (b), the 11 Skyrme parameter sets yield the neutron Fermi level 
$\varepsilon_F \sim -1.9 - -3.9$ MeV, too deep for a halo nucleus. The SIII interaction 
produces the most loosely bound Fermi level with $\varepsilon_F =-1.89$ MeV. 

All Skyrme interactions we choose produce spherical ground states of $^{22}$C and oblate 
ground states of $^{20}$C with quadrupole deformation $\beta_2 \sim -0.23 - -0.33$. 
The two-neutron separation energy  $S_{2n}$, calculated as the
difference in the HF ground state energies between $^{20}$C and $^{22}$C,
is presented in Fig.~\ref{fig.data}(c). 
The observed $S_{2n}= 0.14 \pm 0.39$ MeV~\cite{Gaudefroy10} is smaller than 
any of those calculated values.
This discrepancy can be partially attributed to
the rotation correction for deformed
$^{20}$C, which requires the beyond-mean-field calculation.

The pairing correlation may play some role in these nuclei. To estimate 
its effect, we perform the Hartree-Fock-Bogoliubov (HFB) calculations with 
three different Skyrme functionals (SIII, SkM$^\ast$, and SLy4), using 
available numerical codes of HFBRAD~\cite{HFBRAD} and HFBTHO~\cite{HFBTHO}.
The adopted pairing energy functional produces the average neutron
pairing gap of 1.245 MeV for $^{120}$Sn. We examine the volume, sureface, 
and mixed types of pairing interactions. For $^{22}$C, the pairing gap is 
calculated to vanish with the volume- and mixed-type pairing. Only the 
surface-type pairing produces the finite pairing gap in the ground state.
All these calculations predict the spherical shape for $^{22}$C. The 
surface pairing reduces the two neutron separation energy of $^{22}$C,
which are calculated with HFBRAD as $S_{2n}=0.59$, 0.78, and 1.03 MeV
for SIII, SkM$^\ast$, and SLy4, respectively.
However, it hardly changes the rms radius.
The largest calculated radius for $^{22}$C is 3.24 fm with the SLy4.
This is still significantly smaller than
the value of $5.4\pm 0.9$ fm suggested by experiment~\cite{Tanaka10}.


\subsection{Adjustment of potential}

\begin{figure}[tb]
\begin{center}
\includegraphics[width=0.375\textwidth,keepaspectratio]{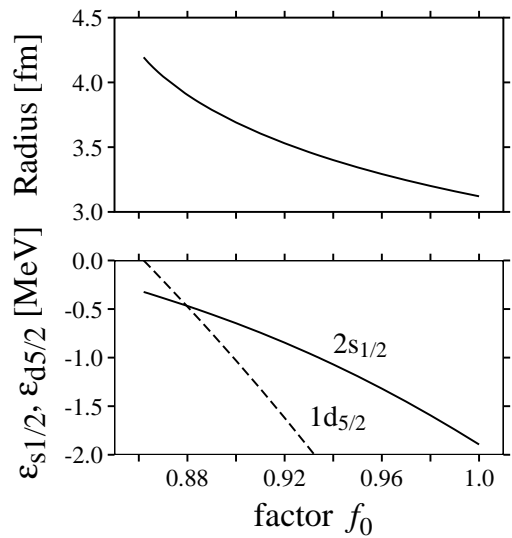}
\caption{
Rms matter radius and single-particle energies of $2s_{1/2}$ (solid) and $1d_{5/2}$ orbits (dashed)
in Skyrme SIII interaction as a function of  $f_0$ value. See text for details.
}
\label{fig.factor}
\end{center}
\end{figure}

The neutron Fermi level is a key ingredient to characterize the neutron drip-line nuclei. 
References~\cite{Tanaka10,Yamashita11,Fortune12,Ershov12} analyzed the large reaction cross 
section $\sigma_R$ by using the $^{20}\mathrm{C}+n+n$ model and concluded that the neutron 
Fermi level should be the $s$ orbit having the single-particle energy $\varepsilon_F \sim 0$ 
MeV to reproduce the large $\sigma_R$ and the corresponding large matter radius. 
In this paper, in order to adjust the matter radius and the neutron Fermi level, we simply 
multiply the parameter $t_0$ in Skyrme interaction by a factor $f_0$, which changes the mean-field central 
potential. Taking smaller value of $f_0<1$, the rms matter radius becomes larger and the 
$2s_{1/2}$ orbit becomes more loosely bound. Figure \ref{fig.factor} shows the case of the 
SIII interaction. As is pointed out in Ref.~\cite{Hamamoto07}, the single-particle energy of 
$1d_{5/2}$ orbit, $\varepsilon_{d5/2}$, is more sensitive to the depth of the central potential, 
rather than $2s_{1/2}$ orbit. This orbit dependence of sensitivity is known to be partially responsible 
for the change of magicity in neutron-rich nuclei. We change the factor $f_0$ while keeping 
$2s_{1/2}$ orbit being the neutron Fermi level, for comparison with results of the 
$^{20}\mathrm{C}+n+n$ model. We find that the modified SIII interaction can produce 
the largest matter radius among the 11 Skyrme interactions we choose. 
The SIII interaction is able to
achieve $\varepsilon_F=-0.50$ MeV on setting $f_0=0.884$, in which the $2s_{1/2}$ and $1d_{5/2}$ 
orbits are almost degenerate ($\varepsilon_{d5/2}=-0.58$ MeV). This modified SIII interaction 
with $f_0=0.884$ yields large $\sigma_R$ comparable to the experimental value, which will be 
discussed in the next subsection. Hereafter we use the modified SIII interaction unless otherwise 
specified. When $f_0=0.884$, the rms radius of $2s_{1/2}$ orbit becomes 7.20 fm and the proton 
and neutron radii are 2.78 and 4.23 fm, respectively. The rms matter radius 3.89 fm is larger 
than the predicted value of Ref.~\cite{Horiuchi06} but smaller than the experiential value~\cite{Tanaka10} 
estimated from $\sigma_R$ using a three-body model.

It should be noted that the $S_{2n}$ value decreases monotonically 
as the mean-field central potential becomes shallow, and turns out to be negative 
when $f_0=0.884$. Therefore, if we set $\varepsilon_F=-0.50$ MeV, $^{22}$C is 
unbound with respect to $^{20}$C due to the deformation in the present calculation. 
The pairing correlation may improve this undesirable situation of $^{22}$C. 
Constructing a new parameter set suitable for describing very neutron-rich nuclei is important but it is beyond the scope of the present work.

\subsection{Total reaction cross section: Glauber model analysis}

By calculating total $\sigma_R$ for nucleus-nucleus collision, 
we test the validity of the modified interaction. 
A high-energy collision is described in the Glauber formalism~\cite{Glauber}.
The $\sigma_R$ is calculated by
\begin{eqnarray}
\sigma_R = \int\! d\bm{b} \, \left( 1-|{\rm e}^{i\chi(\bm{b})}|^2\right),
\end{eqnarray}
where $\chi(\bm{b})$ is a phase shift function describing 
the  collision and the integration is done over 
an impact parameter $\bm{b}$ between the projectile and the target.
Here we use an optical limit approximation (OLA) which
offers a simple expression that only requires
one-body density distributions of the projectile, $\rho_P (\bm{r}^P)$,
and target, $\rho_T (\bm{r}^T)$.
In the OLA, the phase shift function is expressed by
\begin{eqnarray}
&& \mathrm{e}^{ i \chi_\mathrm{OLA}(\bm{b})} =\exp \bigg[-\iint \! d\bm{r}^Pd\bm{r}^T \rho_P (\bm{r}^P)\rho_T (\bm{r}^T) \nonumber\\
&& \phantom{\mathrm{e}^{ i \chi_\mathrm{OLA}(\bm{b})} =\exp \Bigg[-\iint}\quad \times\Gamma_{NN}(\bm{s}^P \- \bm{s}^T \+ \bm{b})\bigg],
\end{eqnarray}
where $\bm{s}^P (\bm{s}^T)$ is the transverse component 
of the projectile (target) coordinate
and $\Gamma_{NN}$ is the nucleon-nucleon profile function whose parameters
are fitted to reproduce the nucleon-nucleon collisions, and
thus the model has no {\it ad hoc} parameter.
The parameter sets used here are listed in Ref.~\cite{Ibrahim08}.
The OLA ignores some higher multiple scattering effects
and usually overestimates slightly the $\sigma_R$~\cite{Horiuchi07}.
We employ another expression,
called nucleon-target formalism in the Glauber model (NTG),
which includes the multiple scattering effects, 
but requires the same input as the OLA~\cite{Ibrahim00}.
The power of this formalism is confirmed in systematic analysis
for carbon~\cite{Horiuchi06, Horiuchi07} and oxygen~\cite{Ibrahim09} 
isotopes as well as light neutron rich nuclei 
with $Z=10-16$~\cite{Horiuchi12}.
We use the OLA for a proton target and the NTG for a carbon target 
in the present analysis.

When the modified SIII interaction is employed,
the calculated $\sigma_R$ on a proton target incident at 40 MeV is 1040 mb.
Though the incident energy is too low for the Glauber approximation to be applied for
a proton target, the $\sigma_R$ appears close 
to the measured cross sections $1338 \pm 274$ mb~\cite{Tanaka10} within the error bar,
whereas the original SIII interaction gives smaller $\sigma_R$, 821 mb.
The modified SIII interaction is more realistic than the original one 
for simulating the ground state of $^{22}$C. The $\sigma_R$ of $^{22}$C on a carbon target
is predicted to be 1480 and 1600 mb incident at 240 and $900 A$MeV, respectively, when 
the modified SIII interaction is employed. The obtained $\sigma_R$s for both proton 
and carbon targets are consistent with those obtained by the three-body 
calculation~\cite{Horiuchi06,Ibrahim08,Horiuchi07}.

\subsection{Low-lying $E1$ strength}

\begin{figure}[tb]
\begin{center}
\includegraphics[width=0.4750\textwidth,keepaspectratio]{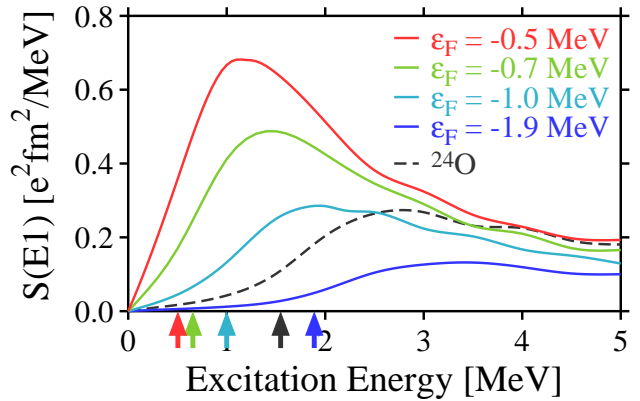}
\caption{(Color online)
Neutron Fermi level dependence of low-lying $E1$ strength. The solid lines from upper 
to lower correspond to the calculations with the neutron Fermi level $\varepsilon_F=-0.5$, 
$-0.7$, $-1.0$, and $-1.9$ MeV, respectively. The dashed line is the result of $^{24}$O with 
$f_0 = 0.884$ corresponding to $\varepsilon_F=0.5$ MeV in $^{22}$C. 
The SIII interaction is used with modification.
}
\label{FermiDep}
\end{center}
\end{figure}

Next we discuss the low-lying $E1$ strength obtained by the self-consistent RPA 
calculation. Figure \ref{FermiDep} demonstrates how the low-lying $E1$ strength develops 
as the neutron Fermi level $\varepsilon_F$ gets closer to zero energy. We calculate the 
low-lying $E1$ strength with the original SIII interaction ($\varepsilon_F=-1.9$ MeV), 
together with the modified interactions that give  $\varepsilon_F= -1.0$ MeV, 
$\varepsilon_F=-0.70$ MeV (corresponding to adjusting the rms matter radius to 3.7 fm~\cite{Horiuchi06}), 
and $\varepsilon_F=-0.50$ MeV. The arrows denote the absolute value 
of $\varepsilon_F$, indicating the threshold of excitation to the continuum states.
The $E1$ strength rises up not from the threshold but from zero excitation energy because 
we calculate the response function at a complex energy with 0.50 MeV of imaginary part. 
In the case of the original SIII interaction, no prominent $E1$ strength is found. As the 
Fermi energy $\varepsilon_F$ approaches zero energy, the low-lying $E1$ strength develops, 
especially at $\varepsilon_F \gesim -1.0$ MeV. This trend is known through the result of 
the three-body cluster model assuming an inert core~\cite{Ershov12}. When 
$\varepsilon_F=-0.50$ MeV, summed $E1$ strengths up to 3.0, 4.0 and 5.0 MeV are 1.38, 
1,65, and 1.85 $e^2$fm$^2$, respectively, which is comparable to the strength of 
$^{11}$Li~\cite{Nakamura06}.

We apply the same factor $f_0 = 0.884$ for $^{24}$O, which corresponds to 
setting $\varepsilon_F=-0.5$ MeV in $^{22}$C. Even though both $^{22}$C and $^{24}$O are neutron drip-line nuclei, 
additional two protons shrink the matter (neutron) radius, 3.89 fm $\to$ 3.42 fm (4.23 fm 
$\to$ 3.64 fm), and shift down the neutron Fermi level by 1 MeV. Consequently, the 
low-lying $E1$ strength of $^{24}$O is not prominent, as shown in Fig.~\ref{FermiDep}.

\subsection{Comparison with giant dipole resonance}

\begin{figure}[tb]
\begin{center}
\includegraphics[width=0.4750\textwidth,keepaspectratio]{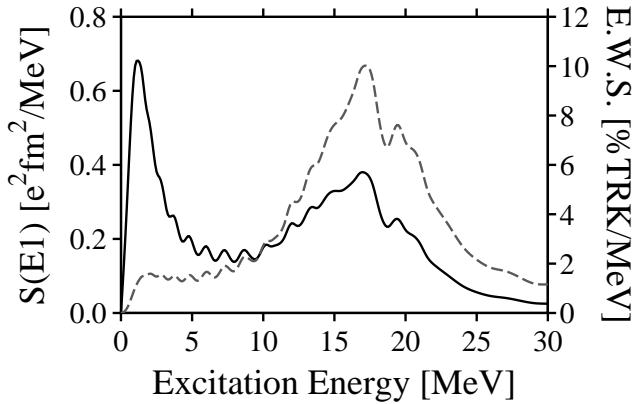}
\caption{$E1$ strength distribution (solid line, left axis) and energy-weighted sum value 
(grayed dashed line, right axis) of $^{22}$C obtained by setting $\varepsilon_F= -0.50$ MeV 
with SIII interaction. 
}
\label{fig.fullE1}
\end{center}
\end{figure}
\begin{figure}[tb]
\begin{center}
\includegraphics[width=0.4990\textwidth,keepaspectratio]{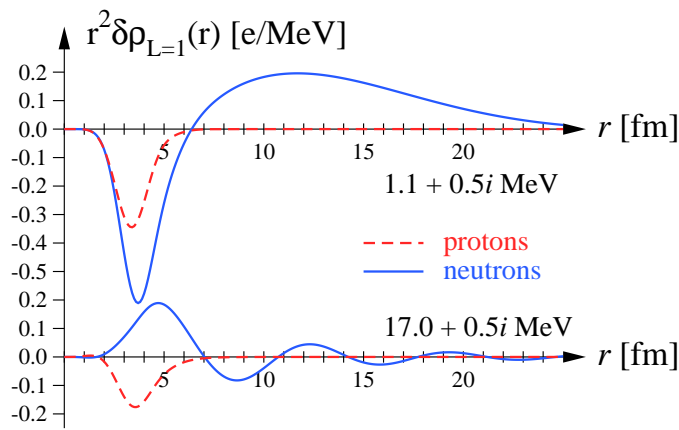}
\caption{(Color online) Dipole transition densities at the peaks of low-lying $E1$ mode (1.1 MeV) and GDR (17.0 MeV).
}
\label{fig.drho}
\end{center}
\end{figure}

An interesting property in drip-line nuclei is the large low-lying $E1$
strength comparable with that of the giant dipole resonance (GDR). 
The full $E1$ strength distribution with $\varepsilon_F= -0.5$ MeV is shown in 
Fig. \ref{fig.fullE1}. Some oscillations appearing at around the excitation energy $\omega 
\sim 5-10$ MeV come from the discretized continuum state and others from proton single 
particle-hole excitations from bound to bound states such as $1p_{3/2} \to 1d_{5/2}$.  
The peak height of the low-lying $E1$ strength in the present calculation
is higher than that of the GDR. This is very unusual. The peak height of the 
observed low-lying $E1$ strength (PDR) in heavy neutron-rich nuclei such as 
$^{132}$Sn~\cite{Adrich05} is always lower than half of the height of the GDR.
In fact, the calculated low-lying $E1$ strength carries about 1/3 of the total $E1$ strength.
Though the peak height depends on the smearing width $\gamma$, we confirm that the peak 
of low-lying $E1$ strength is higher than the GDR when $\gamma \lesim 3$ MeV. 
Our calculation demonstrates that the ``pygmy'' dipole resonance could be a 
``giant'' low-lying dipole resonance in neutron drip-line nuclei. 
The energy-weighted sum of $E1$ strength is also plotted in Fig.~\ref{fig.fullE1}, but 
its discussion will be given later in Sec. \ref{subsec:sumrule}.

In Fig.~\ref{fig.drho}, we plot the dipole transition densities
\begin{eqnarray}
r^2 \delta\rho_{L=1}(r) = r^2 \int d\Omega \,\, Y_1(\Omega) \delta\rho(\vr)
\label{drhoL1}
\end{eqnarray}
of protons and neutrons at the peaks of the low-lying $E1$ (1.1 MeV) and the GDR (17.0 MeV). 
The GDR transition densities display out-of-phase densities 
between protons and neutrons, which is typical for the GDR.
Because of the halo structure in the ground state,
the neutron transition density
of the GDR has an oscillating extended tail.
For the low-lying $E1$ case, the transition densities look like the oscillation 
of the outer neutron against the inner core, namely, the protons and neutrons inside the 
core nucleus oscillates in phase and only those neutrons residing far outside the core 
move out of phase against the inner core, similar to the classical picture
of PDR~\cite{Suzuki90}.
Due to the loosely bound Fermi level with $\varepsilon_F= -0.50$ MeV, the neutron transition 
density has a quite long tail spreading to $r \sim 25$ fm, which indicates excitations 
from $2s_{1/2}$ orbit to the low-energy continuum states.

\subsection{Cluster sum rule value}
\label{subsec:sumrule}

The sum rule is useful for a qualitative estimation of the contribution of the low-lying $E1$ strength.
The energy-weighted sum rule is given by
\begin{eqnarray}
S_\mathrm{TRK} = \frac{9 e^2}{4\pi} \frac{\hbar^2}{2m} \frac{NZ}{A} .
\end{eqnarray}
This is known as the classical Thomas-Reiche-Kuhn (TRK) sum rule. The calculated low-lying 
$E1$ strength carries a sizable contribution despite the quite small excitation energy. 
As shown in Fig.~\ref{fig.fullE1}, the low-lying $E1$ strength distribution exhausts 
6.2, 11.0 and 15.4 \% of the TRK sum rule when the strength is accumulated up to the 
excitation energy 5, 8 and 10 MeV, respectively. This is comparable to $^6$He~\cite{Aumann99} 
and $^{11}$Li \cite{Zinser97}, and is larger than those of the observed PDRs in heavier nuclei.

The low-lying $E1$ strength in light nuclei such as He and Be isotopes is often analyzed with use 
of the cluster model. Suppose that $^{22}$C has a cluster-like structure, 
i.e., two neutrons coupled to the $^{20}$C core. Then 
the energy-weighted cluster sum rule is evaluated as~\cite{Alhassid82,Sagawa90}
\begin{eqnarray}
S_\mathrm{clus} = \frac{9 e^2}{4\pi} \frac{\hbar^2}{2m} \frac{2 Z^2}{A \left( A - 2\right)} .
\end{eqnarray}
The cumulative energy-weighted sum value,
$S(E_c)=\int^{E_c} E' S(E';E1) dE'$, exceeds $S_\mathrm{clus}$ at $E_c=3.3$ MeV in the case of  
$\varepsilon_F= -0.50$ MeV. Even for $\varepsilon_F= -0.70$ MeV or $-1.0$ MeV, it exceeds $S_\mathrm{clus}$
at $E_c=3.9$ MeV and 5.0 MeV, respectively. This means that the three-body model with the $^{20}$C 
core and two neutrons is not supported in the present approach at least for  
$\varepsilon_F \gesim -1.0$ MeV. Exceeding the cluster sum rule indicates 
some other contributions coming from a core excitation, such as excitations from $1d_{5/2}$ orbit. 
Due to the core excitation, the non-energy-weighted cluster sum 
rule in fact does not work for the investigation of the calculated low-lying $E1$ strength. 
In the following, we show that a simple picture of the PDR, two valence neutrons oscillating 
against the $^{20}$C core, is not fully supported by the present calculation.
Contributions of the core excitation are present and discussed in the Sec.~\ref{sec:d5/2}

\subsection{Core excitation: Role of $d_{5/2}$ state}
\label{sec:d5/2}

\begin{figure}[tb]
\begin{center}
\includegraphics[width=0.4990\textwidth,keepaspectratio]{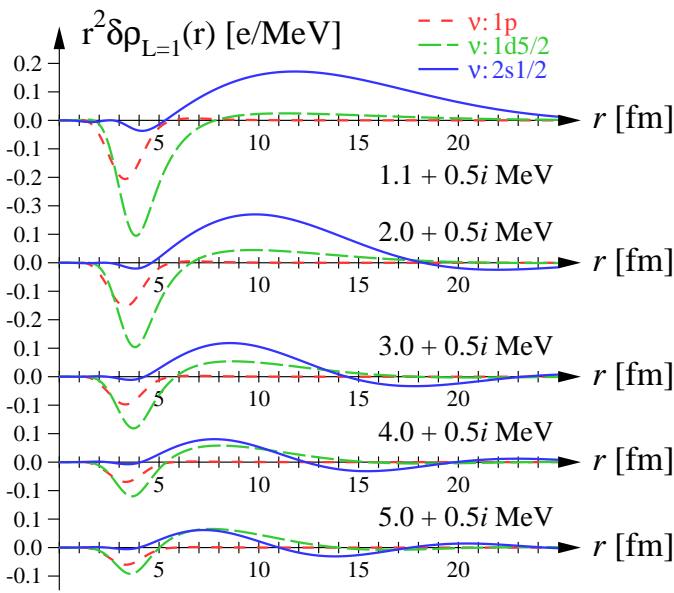}
\caption{(Color online)
Neutron transition densities decomposed to the occupied orbits, 
$1p$ (red), $1d_{5/2}$ (green), and $2s_{1/2}$ (blue) orbits, as a function of radius. 
Omitted transition density from $1s_{1/2}$ orbit is negligible. 
}
\label{fig.drho2}
\end{center}
\end{figure}

As is mentioned above, the neutron excitations from $1d_{5/2}$ orbit may contribute 
to the low-lying $E1$ strength. Moreover, the calculated low-lying $E1$ strength 
has a large tail at excitation energy $\omega \gesim$ 2 MeV (See Fig. \ref{FermiDep}), 
in contrast to those of the three-body calculation~\cite{Ershov12}. This difference 
supports some contributions of $1d_{5/2}$ orbit. Figure \ref{fig.drho2} demonstrates 
that the excitations from $1d_{5/2}$ orbit play a role of enhancing the $E1$ strength. 
We plot the neutron transition densities at a peak of the $E1$ strength, 1.1 MeV, and 
excitation energies from 2.0 MeV to 5.0 MeV with a spacing of 1.0 MeV. The transition 
densities are decomposed to the occupied orbits, namely, the decomposed ones are 
calculated by Eqs.~(\ref{drho}) and (\ref{drhoL1}) but the index $i$ runs over only the 
corresponding orbits and the sum of them is equal to the transition density in 
Fig.~\ref{fig.drho}. At the peak position, the decomposed transition 
densities are divided to that of $2s_{1/2}$ orbit 
and the others. The transition density of $2s_{1/2}$ orbit has a long tail up to 
$r\sim 25$ fm, describing excitations to the low-energy continuum state. The other 
transition densities (including proton transition densities) have the sign opposite to the 
$2s_{1/2}$ transition density and their contributions are small at $r \gesim 5$ fm, 
suggesting a recoil of the $^{20}$C core. Therefore, the transition densities 
at 1.1 MeV show that two neutrons are excited to continuum state and go away from 
the remaining core. Our RPA calculation produces results similar  
to the three-body cluster model at the peak position~\cite{Ershov12}. 
Nevertheless, it is seen even in 1.1 MeV state that a part of neutrons in $1d_{5/2}$ orbit is also excited to the continuum 
state. As excitation energy increases,  the $2s_{1/2}$ transition density 
gradually becomes small, whereas the contribution of $1d_{5/2} \to$ continuum states 
develops and eventually become comparable to that of $2s_{1/2}$ orbit at 5.0 
MeV excitation energy. This $1d_{5/2}$ contribution lifts the $E1$ strength at $E \gesim$ 2 MeV 
and produces the long tail of the low-lying $E1$ strength.

\begin{figure}[tb]
\begin{center}
\includegraphics[width=0.4750\textwidth,keepaspectratio]{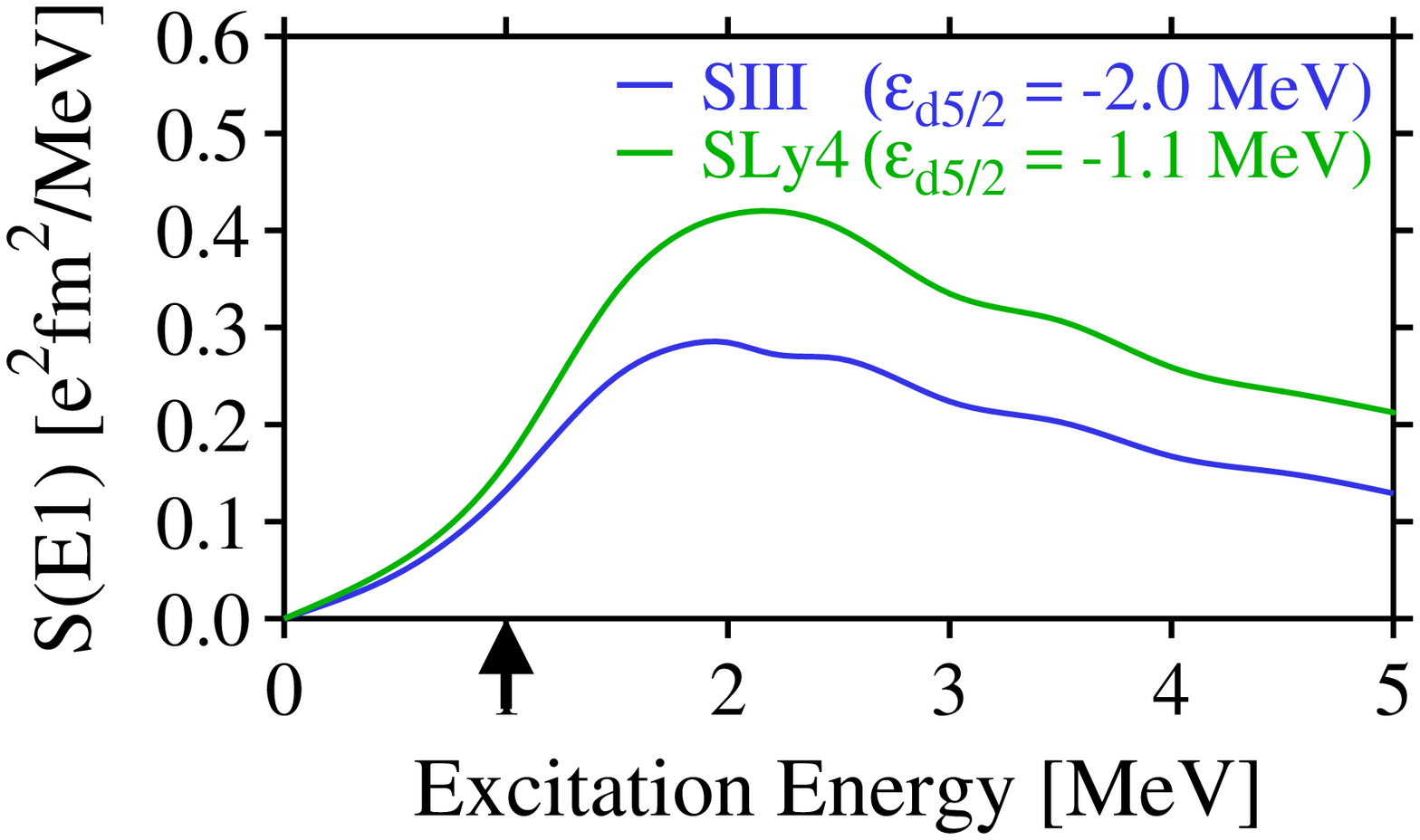}
\includegraphics[width=0.4990\textwidth,keepaspectratio]{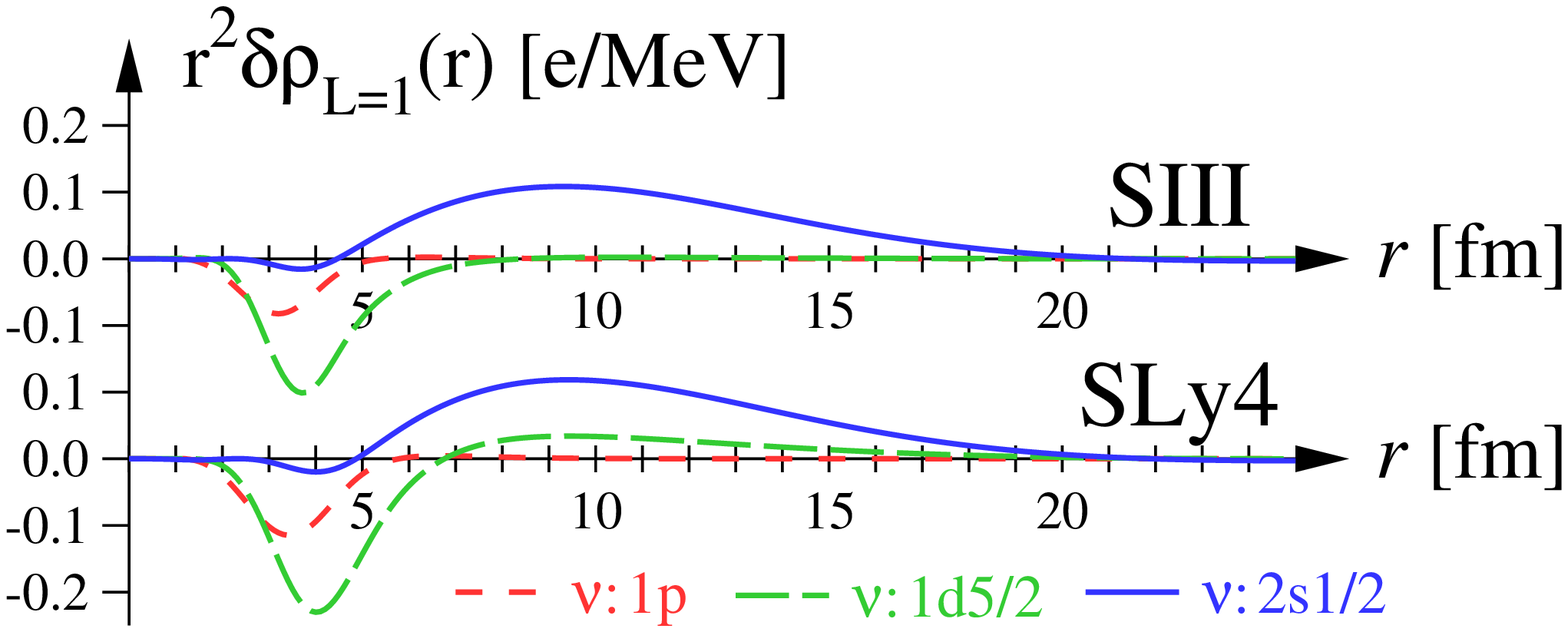}
\caption{(Color online) Comparison of low-lying $E1$ strength (upper) and transition densities (lower) at peak position
calculated with SIII and SLy4 interactions that are modified to set $\varepsilon_F=-1.0$ MeV.
}
\label{SIII.SLy4}
\end{center}
\end{figure}

Comparison of low-lying $E1$ strengths calculated with the same $\varepsilon_F$ but different interactions 
shows more clearly the role of the $1d_{5/2}$ orbit because the low-lying $E1$ strength distribution in 
neutron drip-line nuclei is not sensitive to the interaction itself used in calculation, but sensitive 
to the single-particle properties near the Fermi level. The upper panel of Fig.~\ref{SIII.SLy4} shows the  
low-lying $E1$ strength obtained by setting $\varepsilon_F =-1.0$ MeV with the SIII and SLy4 interactions. 
The SLy4 (SIII) interaction with $\varepsilon_F =-1.0$ 
MeV yields $\varepsilon_{d5/2}=-1.1$ MeV ($-2.0$ MeV). While the peak position of the low-lying $E1$ 
strength is almost the same due to the same $\varepsilon_F$, the $E1$ strength with SLy4 is larger 
than that with SIII. This originates from  the $1d_{5/2}$ orbit, as shown in the 
lower panel of Fig.~\ref{SIII.SLy4} which compares the decomposed transition densities at the 
peak positions. The $2s_{1/2}$ transition densities are quite similar to each other but clear 
difference is seen in the $1d_{5/2}$ transition densities at $r\sim 7-20$ fm. Such difference, 
though it is apparently small, enhances the $E1$ strength by a factor $\sim 1.5$. This indicates 
the importance of the core excitation or explicit treatment of the $1d_{5/2}$ orbit in $^{22}$C.

\subsection{Validity of calculation}

\begin{figure}[tb]
\begin{center}
\includegraphics[width=0.4250\textwidth,keepaspectratio]{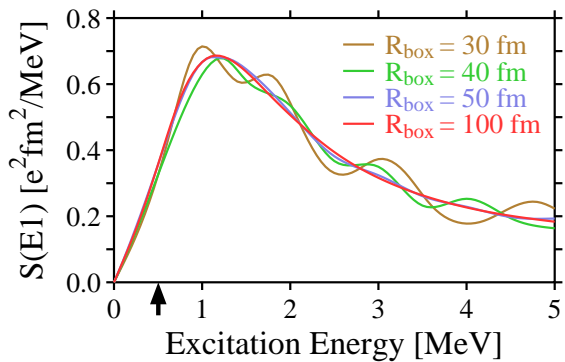}
\caption{(Color online)
Box size dependence of the calculated $E1$ strength with $\varepsilon_F=-0.50$ MeV.
}
\label{fig.Rbox}
\end{center}
\end{figure}

Here we comment on the validity and accuracy of our calculation for the low-lying $E1$ excitation 
in drip-line nuclei. Since the loosely bound $2s_{1/2}$ orbit with $\varepsilon_{s1/2}=-0.50$ MeV 
is spatially quite spread and couples with the continuum state by a small excitation energy, 
a large calculation space is needed to describe the low-lying $E1$ strength properly. Figure \ref{fig.Rbox} 
shows the box size dependence of the calculated low-lying $E1$ strength with $\gamma = 1$ MeV for 
$R_\mathrm{box}=$ 30, 40, 50, and 100 fm. The summed $E1$ strength below 5 MeV is not 
sensitive to $R_\mathrm{box}$; 1.855, 1.809, 1.851, and 1.849 $e^2\mathrm{fm}^2$ for $R_\mathrm{box}=$ 
30, 40, 50, and 100 fm, respectively. While the $E1$ strength distributions with $R_\mathrm{box}=$ 
30 and 40 fm have oscillations stemming from discretized continuum states, the result with 
$R_\mathrm{box}=$ 50 fm shows practically no oscillation and agrees well with $R_\mathrm{box}=$ 100 fm. 
Difference between the calculated $E1$ strengths with $R_\mathrm{box}=$ 50 and 100 fm is less than 0.5 
\% except for very low-energy region, $\omega \lesim 0.5$ MeV. Thus, the box size $R_\mathrm{box}=$ 
50 fm we employ in this paper is large enough for quantitative study of the low-lying $E1$ mode with 
$\gamma = 1$ MeV. 

Next, let us check the validity of the RPA and the mixture of the spurious state.
It is known that the RPA becomes unreliable for low-lying states with very high 
collectivity, because the RPA assumes the small amplitude nature. This can be 
examined by investigating the magnitude of forward and backward amplitudes.
The RPA breaks down if the backward amplitudes become comparable to
the forward ones. For this purpose, we perform the RPA calculation with the 
diagonalization method by using a revised version of the RPA code 
in Ref. \cite{Inakura06} with the same dipole operator (\ref{D_E1}) and for box size of $R_\mathrm{box}=40$ fm. 
The spurious state appears at excitation energy 0.15 MeV.
For this state, the squared modulus of the backward amplitude, 
$\left| Y \right|^2 \equiv  \sum_i \int d\vr\, \left|Y_i(\vr)\right|^2$, 
is 15.0. This means the forward amplitude, $|X|^2=16.0$.
In contrast, the lowest-energy physical 
state appears at 0.75 MeV and its $\left| Y \right|^2$ is 0.042.
Therefore, the present RPA calculation does not break down 
for the low-lying $E1$ modes.

We calculate isoscalar dipole strength (i.e., spurious component) and confirm that 
the spurious state is well separated from the physical states. The isoscalar dipole 
strength is calculated by Eq.~(\ref{dB/dw}) but for an isoscalar dipole 
operator $D_\mathrm{IS}=e \sum_{i=1}^A r_i Y_1(\Omega)$. 
The isoscalar dipole strength for the low-lying peak is satisfactorily small, 
$\left| \langle n | D_\mathrm{IS} | 0 \rangle \right|^2 < 2 \times 10^{-3}$ e$^2$fm$^2$, 
which is about 1 \% of the corresponding $E1$ strength. Therefore, 
numerical calculation of the spurious state is not serious in the present calculation.

It is worthy to note that the low-lying $E1$ strengths smeared by Lorentzians 
(\ref{FAM_smearing}) with $\gamma = 1$ MeV is underestimated compared with those 
calculated by the diagonalization method which corresponds to a limit of 
$\gamma \to 0$ MeV. For example, summations of the $E1$ strengths below 5 MeV, 
calculated by the diagonalization method and the response function with 
$\gamma = 1$ MeV, are 2.366 and 1.809 $e^2\mathrm{fm}^2$, 
respectively, for $R_\mathrm{box}=$ 40 fm. Such underestimation is noticeable 
for low-lying $E1$ strengths but not for the GDR region. Small $\gamma$ is 
obviously better but requires large calculation space for obtaining converged 
$E1$ strength distribution. Thus we employ the $\gamma = 1$ MeV in this paper. 
In addition, we do not take into account
the pairing correlation in the present calculations. 
The pairing correlation does not change the qualitative nature of the $E1$
strength distribution \cite{Ebata10,Ebata11}, but, may enhance the
low-energy $E1$ strength.
Therefore, the calculated low-lying $E1$ strength may become even larger
in more realistic calculations.

\section{Conclusions}
\label{sec.Conclusion}

We have studied the ground state properties and the low-lying $E1$ strength of 
$^{22}$C using the mean-field approach which does not assume the $^{20}$C core. 
Since the original Skyrme interactions we chose do not give a consistent description 
of the observed ground state properties such as the nuclear size, 
we adjusted the central part of the Skyrme potential. 
When we set the neutron Fermi level $\varepsilon_F \gesim -1.0$ MeV, 
the obtained total reaction cross section reasonably agrees with the measured value.
With $\varepsilon_F \gesim -0.5$ MeV, the calculation predicts the low-lying $E1$ strength
comparable with that of the giant dipole resonance. The energy weighted cluster sum rule 
assuming a $^{20}$C core is tested.  The cumulative $E1$ strength exceeds the sum rule 
at very low energy due to the contribution of the ``core'' excitation. Such large 
low-lying $E1$ strength consists mainly of the excitations from $2s_{1/2}$ and $1d_{5/2}$ 
orbits to the continuum states. As the excitation energy increases, the contribution of 
$1d_{5/2}$ orbit to the low-lying $E1$ strength develops and could become comparable to 
that of $2s_{1/2}$ orbit. A precise measurement of the low-lying $E1$ strength and a 
careful analysis of the neutron orbits, e.g., the momentum distribution of $^{20}$C 
fragment of $^{22}$C breakup~\cite{Kobayashi12}, are desired to clarify the role of the 
core excitation in $^{22}$C.

\section*{Acknowledgments}

This work was supported in part by JSPS KAKENHI Grant numbers (25800121, 24540261, 25287065, 24105006).
It is also supported by the HPCI System Research project (Project ID:hp120192).


\begin{thebibliography}{99}

\bibitem{Horiuchi06} 
W. Horiuchi and Y. Suzuki, 
Phys. Rev. C {\bf 74}, 034311 (2006).

\bibitem{Tanaka10} 
K. Tanaka {\it et al.}, 
Phys. Rev. Lett. {\bf 104}, 062701 (2010).

\bibitem{Gaudefroy10} 
L. Gaudefroy {\it et al.}, 
Phys. Rev. Lett. {\bf 109}, 202503 (2012).

\bibitem{Yamashita11}
M.T. Yamashita, R.S. Marques de Carvalho, T. Frederico, Lauro Tomio, 
Phys. Lett. B {\bf 697} (2011) 90-93.

\bibitem{Fortune12}
H.T. Fortune and R. Sherr, 
Phys. Rev. C {\bf 85}, 027303 (2012).

\bibitem{Ershov12}
S.N. Ershov, J.S. Vaagen, M.V. Zhukov, 
Phys. Rev. C {\bf 86}, 034331 (2012).

\bibitem{Carbone10} 
A. Carbone {\it et al.}, Phys. Rev. C {\bf 81}, 041301(R) (2010), and references therein.

\bibitem{Otsuka02}
T. Otsuka, Y. Utsuno, R. Fujimoto, B.A. Brown, M. Honma, and T. Mizusaki, 
Eur. Phys. J. {\bf A 15}, 151-155 (2002).

\bibitem{Hamamoto07}
I. Hamamoto, 
Phys. Rev. C {\bf 76}, 054319 (2007).

\bibitem{Vautherin72} 
D. Vautherin and D.M. Brink, 
Phys. Rev. C {\bf 5}, 626 (1972).

\bibitem{Nakatsukasa05} 
T. Nakatsukasa and K. Yabana, 
Phys. Rev. C {\bf 71}, 024301 (2005).

\bibitem{Flocard78}
H. Flocard, S.E. Koonin, and M.S. Weiss, 
Phys. Rev. C {\bf 17}, 1682 (1978).

\bibitem{Davies80} 
K.T.R. Davies, H. Flocard, S. Krieger, and M.S. Weiss, 
Nucl. Phys. A {\bf 342}, 111 (1980).

\bibitem{Ring-Schuck}
P. Ring and P. Schuck, {\it The Nuclear Many-Body Problem}, (Springer-Verlag, 1980).

\bibitem{Eisenstat}
S.C. Eisenstat, H.C. Elman, and M.H. Schultz, SIAM J. Numer, Anal, {\bf 20}, 345-357  (1983). 

\bibitem{Nakatsukasa07} 
T. Nakatsukasa, T. Inakura, and K. Yabana, 
Phys. Rev. C {\bf 76}, 024318 (2007).

\bibitem{Inakura09} 
T. Inakura, T. Nakatsukasa, and K. Yabana, 
Phys. Rev. C {\bf 80}, 044301 (2009).

\bibitem{Inakura11}
T. Inakura, T. Nakatsukasa, and K. Yabana, 
Phys. Rev. C {\bf 84}, 021302(R) (2011).

\bibitem{Avogadro11}
P. Avogadro and T. Nakatsukasa, 
Phys. Rev. C {\bf 84}, 014314 (2011).

\bibitem{Stoitsov11}
M. Stoitsov, M. Kortelainen, T. Nakatsukasa, C. Losa, and W. Nazarewicz, 
Phys. Rev. C {\bf 84}, 041305(R) (2011).

\bibitem{Avogadro13}
P. Avogadro and T. Nakatsukasa, 
Phys. Rev. C {\bf 87}, 014331 (2013).

\bibitem{Liang13}
H. Z. Liang, T. Nakatsukasa,  Z. Niu, and J. Meng,
Phys. Rev. C {\bf 87}, 054310 (2013).

\bibitem{Hinohara13}
N. Hinohara, M. Kortelainen, and W. Nazarewicz,
Phys. Rev. C {\bf 87}, 064309 (2013).

\bibitem{SIII} 
M. Beiner, H. Flocard, Nguyen van Giai, and P. Quentin, 
Nucl. Phys. A {\bf 238}, 29 (1975).

\bibitem{SLy4} 
E. Chabanat, P. Bonche, P. Haensel, J. Mayer, and R. Schaeffer, 
Nucl. Phys. A {\bf 627}, 231 (1998).

\bibitem{SGII} 
Nguyen Van Giai, and H. Sagawa, 
Phys. Lett. B {\bf 106}, 379 (1981).

\bibitem{SkM*} 
J. Bartel, P. Quentin, M. Brack, C. Guet, and H.B. H\r{a}kansson, 
Nucl. Phys. A {\bf 386}, 79 (1982).

\bibitem{SVmin} 
P. Kl\"upfel, P.-G. Reinhard, T.J. B\"urvenich, and J.A. Maruhn, 
Phys. Rev. C 79, 034310 (2009)

\bibitem{UNEDF} 
M. Kortelainen, J. McDonnell, W. Nazarewicz, P.-G. Reinhard, J. Sarich, N. Schunck, M.V. Stoitsov, and S.M. Wild, 
Phys. Rev. C {\bf 85}, 024304 (2012)

\bibitem{SkI} 
P.-G. Reinhard and H. Flocard, 
Nucl. Phys. A {\bf 584}, 467 (1995).

\bibitem{HFBRAD}
K. Bennaceur and J. Dobaczewski, Computer Physics Communications {\bf 168} (2005) 96-122.

\bibitem{HFBTHO}
M.V. Stoitsov, N. Schunck, M. Kortelainen, N. Michel, H. Nam, E. Olsen, J. Sarich, and S. Wild, 
Computer Physics Communications {\bf 184} (2013) 1592-1604.


\bibitem{Glauber}
R. J. Glauber, in Lecture in Theoretical Physics, edited by W. E. Brittin and L. G. Dunham, 
Vol. 1 (Interscience, New York, 1959), p. 315.

\bibitem{Ibrahim08}
B. Abu-Ibrahim, W. Horiuchi, A. Kohama, and Y. Suzuki,
Phys. Rev. C {\bf 77}, 034607 (2008).

\bibitem{Horiuchi07}
W. Horiuchi, Y. Suzuki, B. Abu-Ibrahim, and A. Kohama,
Phys. Phys. C {\bf 75}, 044607 (2007).

\bibitem{Ibrahim00}
B. Abu-Ibrahim and Y. Suzuki, 
Phys. Rev. C {\bf 61}, 051601(R) (2000).

\bibitem{Ibrahim09}
B. Abu-Ibrahim, S. Iwasaki, W. Horiuchi, A. Kohama, and Y. Suzuki,
J. Phys. Soc. Jpn. {\bf 78}, 044201 (2009).

\bibitem{Horiuchi12}
W. Horiuchi, T. Inakura, T. Nakatsukasa, and Y. Suzuki,
Phys. Rev. C {\bf 86}, 024614 (2012).

\bibitem{Nakamura06}
T. Nakamura {\it et al.}, 
Phys. Rrv. Lett. {\bf 96}, 252502 (2006)

\bibitem{Adrich05}
P. Adrich {\it et al.}, Phys. Rev. Lett. {\bf 95}, 132501 (2005).

\bibitem{Suzuki90}
Y. Suzuki, K. Ikeda, and H. Sato, 
Prog. Theor. Phys. {\bf 83}, 180 (1990).

\bibitem{Aumann99} 
T. Aumann {\it et al.}, 
Phys. Rev. C {\bf 59}, 1252-1262 (1999).

\bibitem{Zinser97} 
M. Zinser {\it et al.}, 
Nucl. Phys. A {\bf 619}, 151 (1997).

\bibitem{Alhassid82}
Y. Alhassid, M. Gai, and G.F. Bertsch, 
Phys. Rev. Lett. {\bf 49}, 1482 (1982).

\bibitem{Sagawa90}
H. Sagawa and M. Honma, 
Phys. Lett. B {\bf 251}, 17 (1990).

\bibitem{Inakura06} 
T. Inakura, H. Imagawa, Y. Hashimoto, S. Mizutori, M. Yamagami K. Matsuyanagi, 
Nucl. Phys. A {\bf 768}, 61 (2006).

\bibitem{Ebata10} 
S. Ebata, T. Nakatsukasa, T. Inakura, K. Yoshida, Y. Hashimoto, and K. Yabana, 
Phys. Rev. C {\bf 82}, 034306 (2010). 

\bibitem{Ebata11} 
S. Ebata, T. Nakatsukasa, and T. Inakura, 
AIP conf. Proc. {\bf 1484}, 427 (2012); 
J. Phys. Conf. Ser. {\bf 381}, 012104 (2012).

\bibitem{Kobayashi12}
N. Kobayashi {\it et al.}, 
Phys. Rev. C {\bf 86}, 054604 (2012).
\end{thebibliography}
\end{document}